\documentclass[reprint, amsmath,amssymb, aip]{revtex4-1}

\usepackage{graphicx}
\usepackage{dcolumn}

\usepackage{bm}
\usepackage{upgreek}
\newcommand{\commentOut}[1]{}

\begin{document}


\title{Excess noise in Al${}_\text{x}$Ga${}_\text{1-x}$As/GaAs based quantum rings}

\author{Christian Riha}
\email[E-mail: ]{riha@physik.hu-berlin.de}
\affiliation{%
Novel Materials Group, Humboldt-Universit\"at zu Berlin, 10099 Berlin, Germany
}%

\author{Sven S. Buchholz}%
\affiliation{%
Novel Materials Group, Humboldt-Universit\"at zu Berlin, 10099 Berlin, Germany
}%

\author{Olivio Chiatti}%
\affiliation{%
Novel Materials Group, Humboldt-Universit\"at zu Berlin, 10099 Berlin, Germany
}%

\author{Andreas D. Wieck}
\affiliation{
 Angewandte Festk\"orperphysik, Ruhr-Universit\"at Bochum, 44780 Bochum, Germany
}%

\author{Dirk Reuter}
\affiliation{%
 Optoelektronische Materialien und Bauelemente, Universit\"at Paderborn, 33098 Paderborn, Germany 
}

\author{Saskia~F.~Fischer}%
\email[E-mail: ]{Saskia.Fischer@physik.hu-berlin.de}
\affiliation{%
Novel Materials Group, Humboldt-Universit\"at zu Berlin, 10099 Berlin, Germany
}%

\date{\today}

\begin{abstract}
Cross-correlated noise measurements are performed in etched Al${}_\text{x}$Ga${}_\text{1-x}$As/GaAs based quantum rings in equilibrium at bath temperature of $T_\text{bath}=4.2\text{ K}$. The measured white noise exceeds the thermal (Johnson-Nyquist) noise  expected from the measured electron temperature $T_\text{e}$ and the electrical resistance $R$. This excess part of the white noise decreases as $T_\text{bath}$ increases and vanishes for $T_\text{bath}\geq 12\text{ K}$. Excess noise is neither observed if one arm of a quantum ring is depleted of electrons nor in 1D-constrictions that have a length and width comparable to the quantum rings. A model is presented that suggests that the excess noise originates from the correlation of noise sources, mediated by phase-coherent propagation of electrons.
\end{abstract}

\maketitle

\section*{I.  Introduction}
In signal processing, noise is considered an unwanted phenomenon, which limits the achievable accuracy of measurement devices. However, noise spectra follow statistics that are specific to the origin of the noise and provide information about the investigated system \cite{motchenbacher1993low}.\\
In 1928 J. B. Johnson and H. Nyquist discovered thermal noise \cite{nyquist1928thermal} and attributed it to the thermal agitation of electrons. Due to its connection to the electron temperature, thermal noise can be employed as primary thermometry that measures the electron temperature independently from the lattice temperature \cite{buchholz2012noise}. The frequency-independent thermal noise is always present in a resistive material as background noise at temperatures above 0 K and is known to  be present in ballistic devices \cite{van1986noise,blanter2000shot,kubo1966fluctuation}.\\
In a circuit, each resistor can be considered as noise source and the total noise is easily derived from the total resistance of the circuit. This holds true if the noise sources are not correlated to each other, as is normally the case for thermal noise due to its random nature. The correlation of noise sources in electronic devices may occur if coupling between the noise sources exists\cite{vasilescu2006electronic,jarrix1997noise, zhang1984correlation,van1986noise}.

Here, we investigate whether the correlation of noise sources is mediated by the phase coherence of electrons in etched quasi one-dimensional (1D) quantum rings at $T_\text{bath}\geq 4.2\text{ K}$. The quantum rings are based on an Al${}_\text{x}$Ga${}_\text{1-x}$As/GaAs heterostructure and the thermal noise measurements are carried out by using the cross-correlation technique \cite{riha2015mode}. Single 1D constrictions, i.e. a quantum point contact (QPC) and a bent and a straight quasi 1D electron waveguide, serve as reference for the quantum ring. We find that if two arms of the ring act as correlated noise sources the measured noise is enhanced.
\section*{II.  Experimental details}
The quantum devices were prepared from the same Al${}_\text{x}$Ga${}_\text{1-x}$As/GaAs heterostructure grown by molecular-beam epitaxy. From Shubnikov-de Haas measurements the electron density is found to be $n=2.07\cdot 10^{11}\text{ cm}^{-2}$ and the electron mobility to be $\mu=2.43\cdot 10^5\text{ cm}^2\text{/Vs}$ at $T=4.2\text{ K}$. From this it can be calculated that the wafer hosts a two-dimensional electron gas (2DEG) with a mean free path of $l\approx 18\text{ }\upmu\text{m}$ and a Fermi wavelength of $\lambda_\text{F}\approx 55\text{ nm}$. The 2DEG is located 110 nm below the surface. The nanostructure was defined by electron-beam lithography and subsequent wet-chemical etching \cite{buchholz2012noise}. The 1D constrictions serve as reference structures among which there is a quantum point contact (QPC) with \mbox{100 nm} width and length. Two further 1D constrictions are a straight and a bent waveguide both with a length of about $3\text{ }\upmu\text{m}$ and a width of about $285\text{ nm}$. These length scales allow for ballistic electron transport through the quantum devices. Each 1D constriction is covered by a global top gate that forms a Ti/Au Schottky contact. Two quantum rings have an asymmetric geometry with a straight and a bent arm. The length and bend of the arms is comparable to the geometry of the reference structures, whereas the width is about $510\text{ nm}$. In one quantum ring each arm is covered by a finger gate and the straight arm hosts a QPC. Scanning electron micrographs (SEM) of the quantum devices are shown in Figs.~2-4.
\begin{figure}[b!]
\centering
\includegraphics[width=0.48\textwidth]{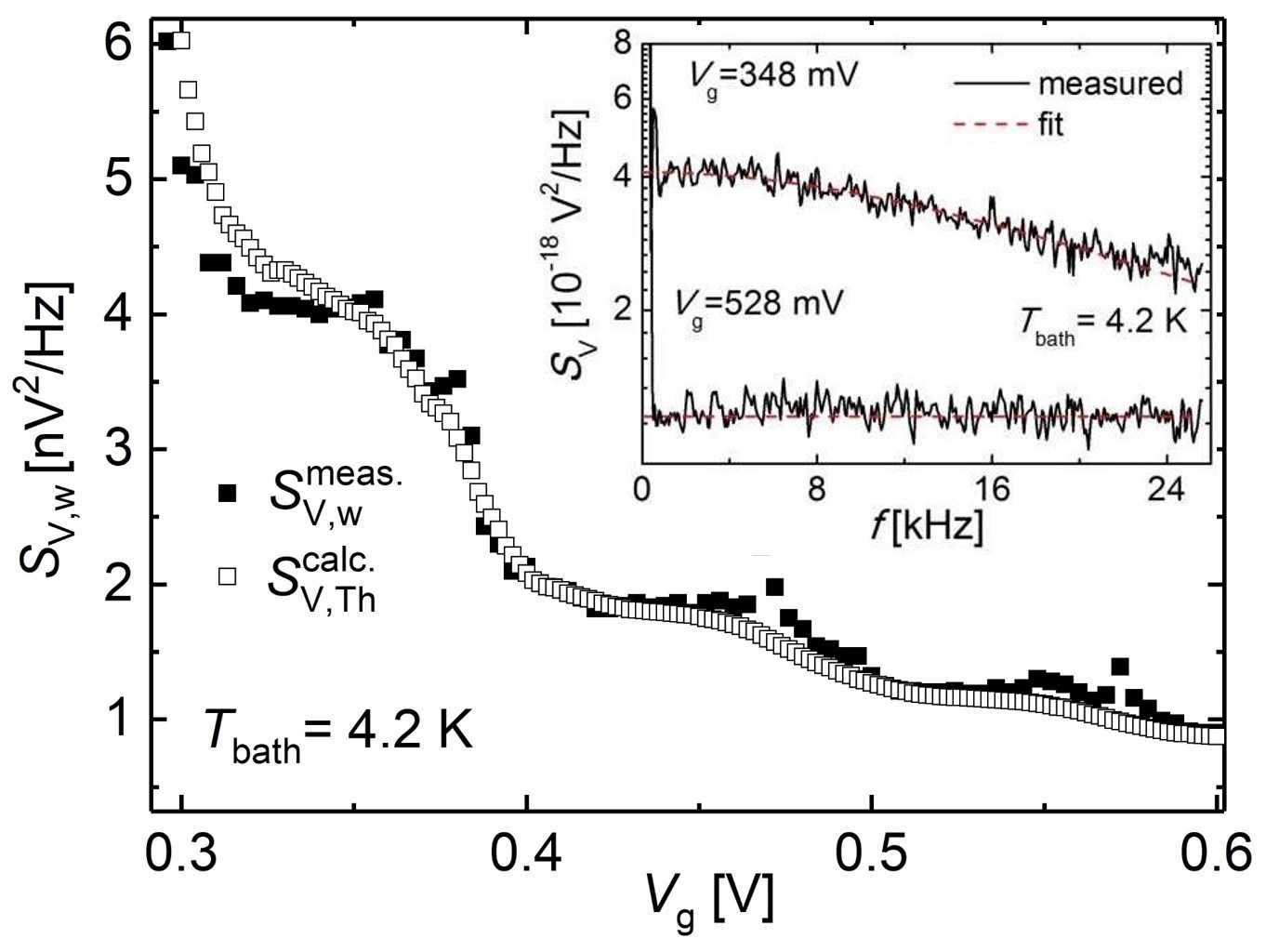}
\caption{Inset: Measured noise spectra of the QPC at gate voltages $V_\text{g}=348\text{ mV}$ (1st plateau) and $V_\text{g}=528\text{ mV}$ (3rd plateau) at $T_\text{bath}=4.2\text{ K}$. The red dashed line indicates the fit according to Eq.~\ref{Tiefpass} (see text). Comparison between measured thermal noise (solid squares) and the theoretically expected thermal noise (open squares) for a quantum point contact.}
\label{1Dcheckoo}
\end{figure}
Cross-correlated noise measurements are carried out at $4.2\text{ K}$ in a frequency range of 1 Hz to 60 kHz with a SR875 spectrum analyzer and two SR5184 ultra low-noise voltage amplifiers with a voltage gain of $10^3$ each. Parasitic capacities $C_\text{par}$ in the measurement setup are
taken into account by fitting the noise spectra with
\begin{equation}
S_\text{V}(f)=\frac{S_\text{V,w}^\text{meas.}}{1+(2\pi f RC_\text{par})^2}
\label{Tiefpass}
\end{equation} 
in order to obtain the white part of the noise $S_\text{V,w}^\text{meas.}$. From the bath temperature $T_\text{bath}=4.2\text{ K}$ and the sample resistance $R$ the expected thermal noise can be calculated with
\begin{equation}
S_\text{V,w}^\text{calc.}=4k_\text{B}T_\text{e}R+2\cdot 4k_\text{B}T_\text{amp}R^2/R_\text{amp}+S_\text{V,l}.
\label{Rauschenbasis}
\end{equation}
The sample resistance is determined by using a lock in amplifier SR830. In Eq.~\ref{Rauschenbasis} the first term represents the Johnson noise \cite{nyquist1928thermal}, where the electron temperature is assumed to be $T_\text{e} = T_\text{bath}$ in the absence of current heating. The second term comes from the ultra low-noise amplifiers with input resistance $R_\text{amp}=5\text{ M}\Omega$ and temperature $T_\text{amp}=300\text{ K}$. Contributions from shot noise are not taken into consideration since the noise measurements are performed in the absence of electric current. The third term represents noise from series resistances. Since the noise measurements are performed in a two-point setup it is important that $R$ is much larger than the series resistance $R_\text{s}$, in order to prevent a significant contribution of $S_\text{V,l}$. 
\section*{III. Experimental results}
\begin{figure}[b!]
\centering
\includegraphics[width=0.48\textwidth]{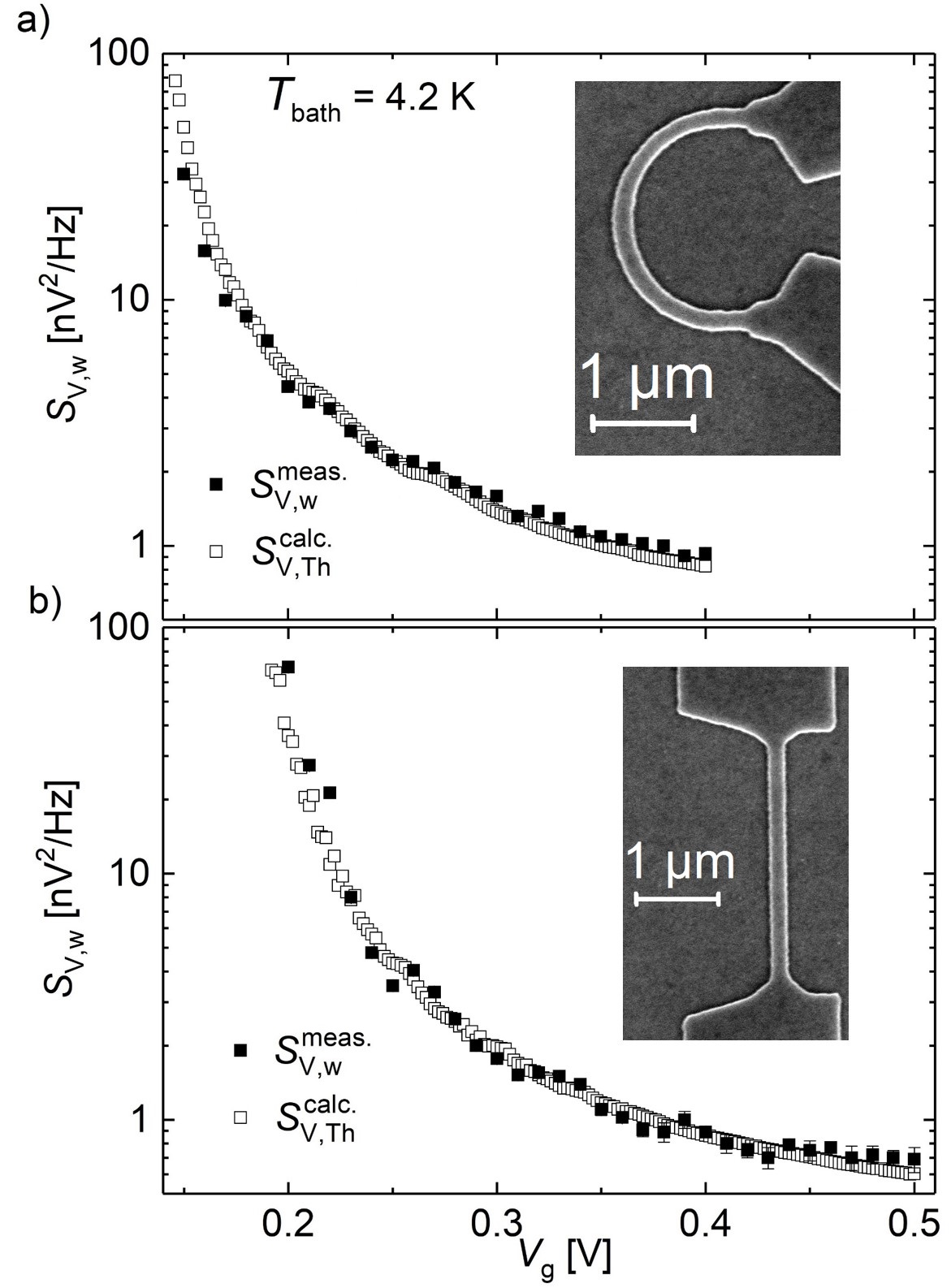}
\caption{Comparison between measured thermal noise (solid squares) and the theoretically expected thermal noise (open squares) for a) a bent 1D constriction and b) a straight 1D constriction  at $T_\text{bath}=4.2\text{ K}$. The insets show SEM images of the devices.}
\label{1Dchecko}
\end{figure}
In a first step the thermal noise in the 1D constrictions is measured. For the QPC the noise measurements are performed for voltages $0.3\text{ V}\leq V_\text{g}\leq 0.6\text{ V}$ applied to the global top gate. These gate voltages correspond to the first three populated subbands of the QPC. From the measured conductance plateaus (not shown here) a series resistance of $R_\text{s}\approx 142\text{ }\Omega$ can be derived, which is small compared to the first three conductance plateaus. In the inset of Fig.~\ref{1Dcheckoo} the measured noise spectra of the QPC are depicted for one and three populated subbands. 
The white part of the noise $S_\text{V,w}$ is extracted by applying Eq.~\ref{Tiefpass} to the noise spectra. In this way $S_\text{V,w}$ is derived for each noise measurement and depicted in Fig.~\ref{1Dcheckoo} (solid squares). The measured data are compared to the expected values of the thermal noise derived from Eq.~\ref{Rauschenbasis} (open squares). The measured values fit the expected values within the measurement uncertainty (size of squares) in the whole measurement range $0.3\text{ V}\leq V_\text{g}\leq 0.6\text{ V}$.  

The same procedure is followed for the bent and the straight 1D constriction at $T_\text{bath}=4.2\text{ K}$. Both devices  are covered by a global top gate. From the measured electrical conductance (not shown here) a series resistance of $R_\text{s}=580\text{ }\Omega$ for the bent waveguide and $R_\text{s}=710\text{ }\Omega$ for the straight waveguide can be derived. The measured white part of the noise $S_\text{V,w}$ and the calculated noise is compared in Fig.~\ref{1Dchecko} for both devices. Again, the measured and the calculated noise fit each other within the measurement uncertainty. 

\begin{figure}[b!]
\centering
\includegraphics[width=0.48\textwidth]{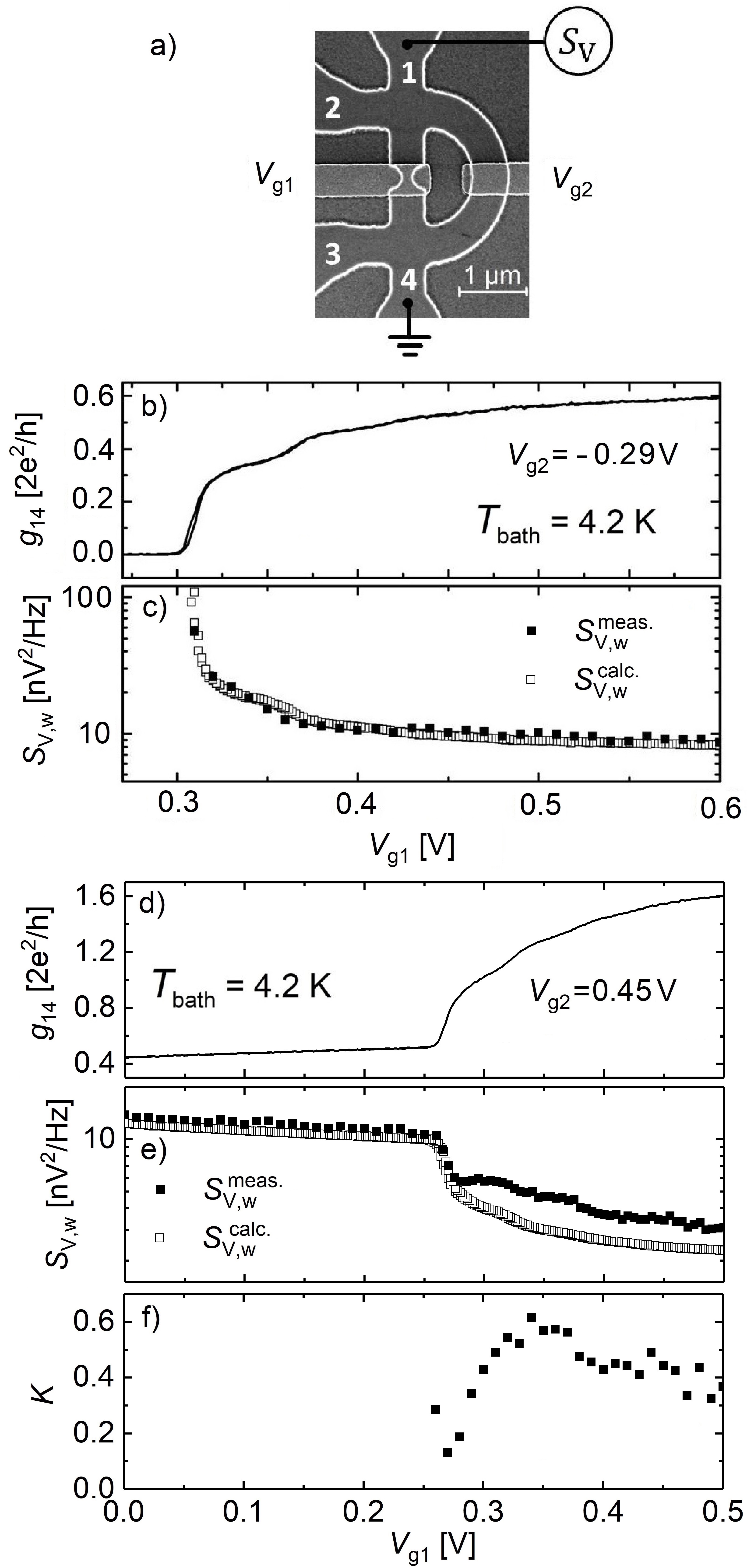}
\caption{Noise measurement across an asymmetric quantum ring at $T_\text{bath}=4.2\text{ K}$. a) SEM image of the device with contacts labelled 1 to 4. Both arms are covered by finger gates and the straight arm hosts a QPC. The gate voltages are labelled $V_\text{g1}$ and $V_\text{g2}$. b) Measured electrical conductance $g_\text{14}$ for $V_\text{g2}=-0.29\text{ V}$. c) Comparision of the measured and the calculated thermal noise for $V_\text{g2}=-0.29\text{ V}$. d) Measured electrical conductance $g_\text{14}$ for $V_\text{g2}=0.45\text{ V}$. e) Comparision of the measured and the calculated thermal noise for $V_\text{g2}=0.45\text{ V}$. f) Calculated values of the correlation coefficient (see text).}
\label{1Dcheck}
\end{figure}

In a second step, the same investigation is performed at $T_\text{bath}=4.2\text{ K}$ for the quantum ring whose arms are each covered by a finger gate. In Fig.~\ref{1Dcheck}~a) a SEM image of the device is depicted with contacts labeled 1 to 4. The gate voltages of the finger gates are labeled $V_\text{g1}$ for the gate that covers the straight arm (with the incorporated QPC) and $V_\text{g2}$ for the gate that covers the bent arm. For $V_\text{g1}>0.31\text{ V}$ and $V_\text{g2}=-0.29\text{ V}$ only the straight arm is electrically conducting. In Fig.~\ref{1Dcheck}~b) the electrical conductance $g_\text{14}$ between contact 1 and contact 4 is shown for $V_\text{g2}=-0.29\text{ V}$ when $V_\text{g1}$ is swept. The measured thermal noise and the noise calculated from $g_\text{14}$ and $T_\text{bath}$ fit each other in the investigated range of $V_\text{g1}$ as depicted in Fig.~\ref{1Dcheck}~c). The same measurement is performed for $V_\text{g2}=0.45\text{ V}$, i.e. when the bent waveguide is electrically conducting. In Fig.~\ref{1Dcheck}~d) the electrical conductance $g_\text{14}$ is depicted for $V_\text{g2}=0.45\text{ V}$ when $V_\text{g1}$ is swept. For $V_\text{g1}<0.27\text{ V}$ only the bent arm of the quantum ring is electrically conducting. In this regime it is found that the measured and the calculated noise deviate by only about 5\% from each other, as depicted in Fig.~\ref{1Dcheck}~e). As long as only one arm is electrically conducting the match between measured and calculated noise is as good as for the 1D reference structures.

In the last step, the gate voltage $V_\text{g1}$ is swept for \mbox{$V_\text{g1}>0.27\text{ V}$} and $V_\text{g2}=0.45\text{ V}$, i.e.  both arms of the ring are electrically conducting. In Fig.~\ref{1Dcheck}~e) it can be seen that the measured noise exceeds the calculated noise by up to 60\%. In the following this is referred to as excess noise: $S_\text{V,w}^\text{excess}=S_\text{V,w}^\text{meas.}-S_\text{V,w}^\text{calc.}$.
\section*{IV. Discussion}
\begin{figure}[t!]
\centering
\includegraphics[width=0.48\textwidth]{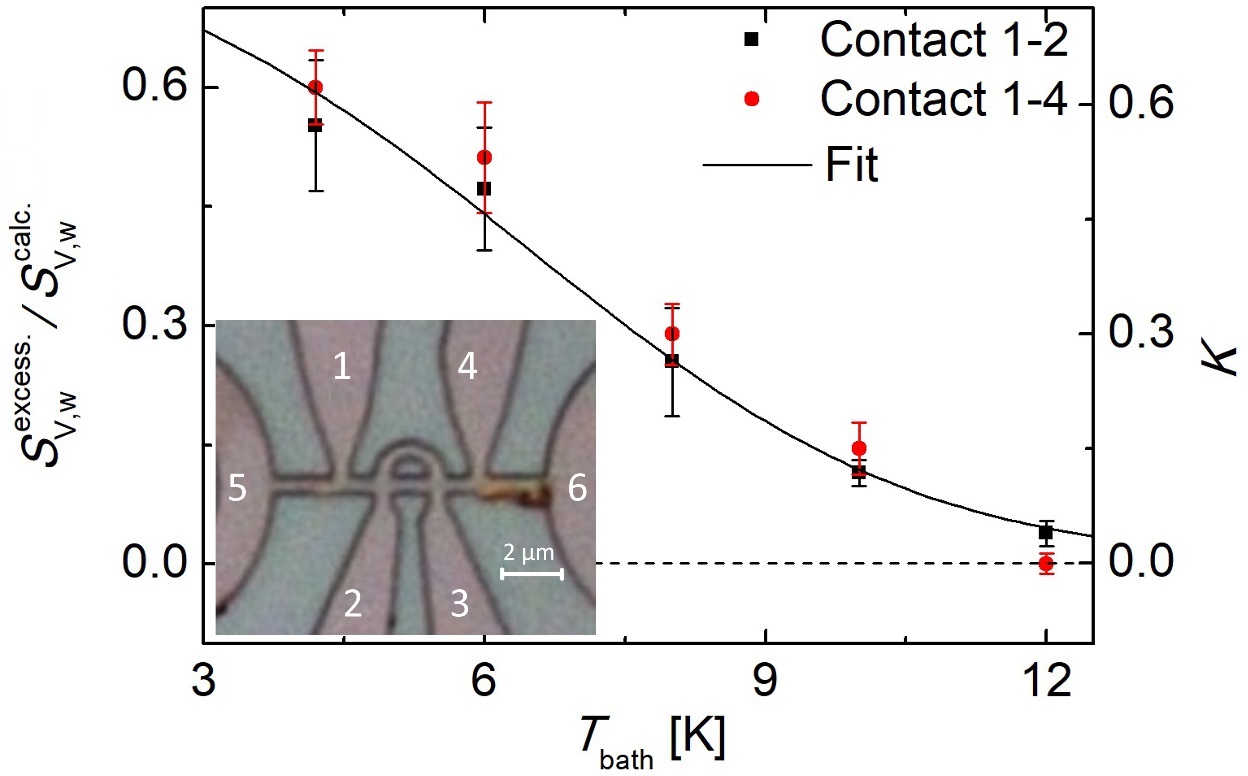}
\caption{Temperature-dependent ratio of the excess noise and the calculated thermal noise of the quantum ring depicted in the inset. The correlation coefficient $K$ is calculated according to Eq.~\ref{corrnoiserationa} (see text). The inset shows an optical micrograph of the device with contacts labeled 1 to 6.}
\label{1Dchec}
\end{figure}
The excess noise in Fig.~\ref{1Dcheck}~e) in the quantum ring is visible for $V_\text{g1}>0.27\text{ V}$, i.e. if both arms are electrically conducting. It is not observed if either one of the arms is depleted of electrons. Since the excess noise is also not observed in the 1D reference structures, it is concluded that the excess noise is a property of a quantum ring when two 'paths' are available to electrons.

In a first approximation, the quantum ring can be understood as a parallel circuit of resistances corresponding to the bent and the straight arm, respectively. Each of these can be considered as source of a thermal noise signal. In a parallel circuit it is more convenient to describe the current noise $S_\text{I}=4k_\text{B}T_\text{e}g$ with $g$ the electrical conductance of the sample. The two parallel resistors $R_1$ and $R_2$ exhibit the noise $S_\text{I,1}$ and $S_\text{I,2}$ that add up to $S_\text{I}=S_\text{I,1}+S_\text{I,2}$ in the absence of correlation between the two noise sources \cite{van1986location}. However, if correlation exists between the noise sources, the resulting total noise reads 
\begin{equation}
S_\text{I,corr}=S_\text{I,1}+S_\text{I,2}+2K\sqrt{S_\text{I,1}S_\text{I,2}}
\label{corrnoise}
\end{equation}
with $K$ the correlation coefficient \cite{vasilescu2006electronic,jarrix1997noise, zhang1984correlation,van1986noise}. If the excess noise is considered to originate from correlation of noise sources in the ring, the ratio $S_\text{V,w}^\text{excess}/S_\text{V,w}^\text{calc.}$ can be expressed as
\begin{equation}
\frac{S_\text{V,w}^\text{excess}}{S_\text{V,w}^\text{calc.}}\equiv\frac{S_\text{I,corr}-S_\text{I,uncorr}}{S_\text{I,uncorr}}=2K\frac{\sqrt{S_\text{I,1}S_\text{I,2}}}{S_\text{I,1}+S_\text{I,2}}.
\label{corrnoiseratio}
\end{equation}
From Eq.~\ref{corrnoiseratio} and the relation $S_\text{I}=4k_\text{B}T_\text{e}g$ the correlation coefficient $K$ can be determined from the measurements
\begin{equation}
K=   \frac{S_\text{V,w}^\text{excess}}{S_\text{V,w}^\text{calc.}}\cdot     \frac{g_1+g_2}{2\sqrt{g_1g_2}}.
\label{corrnoiserationa}
\end{equation}
In our measurements the coefficient can be estimated to range from $K=0.0$ to $K=0.6$, as depicted in Fig.~\ref{1Dcheck}~f). 

Since excess noise is only found in the quantum ring if both arms are conducting, the assumption is that the correlation of noise sources is connected to the phase-coherent propagation of electrons in the ring. In this case $K$ should be related to the visibility $v$ of interference effects such as Aharonov-Bohm oscillations \cite{buchholz2012noise,hansen2001mesoscopic}. The visibility depends on the phase breaking time $\tau_\varphi$ and the traversal time of an electron for going through one arm of the ring $\tau_\text{t}=L/v_\text{F}$ 
according to $v\propto \exp(-\tau_\text{t}/\tau_\varphi)$ and is therefore temperature-dependent \cite{seelig2001charge,hansen2001mesoscopic}. 
The temperature dependence of $K$ was investigated in another quantum ring with a width of $w=500\text{ nm}$. An optical micrograph of the ring is depicted in the inset of Fig.~\ref{1Dchec}. In this ring the excess noise is studied and depicted in Fig.~\ref{1Dchec}. By using Eq.~\ref{corrnoiserationa} and estimating $g_1$ and $g_2$ from four-point measurements, $K$ is estimated. In this sample the thermal length $l_\text{T}=\sqrt{\hbar D/(k_\text{B}T_\text{bath})}\approx 3\text{ }\upmu\text{m}$ is large compared to $w$ in the temperature range $T_\text{bath}= 4.2\text{ K}-12\text{ K}$. This allows to approximate the phase breaking time by the Nyquist scattering time \cite{noguchi1996saturation} $\tau_\text{N}=(\sqrt{E_\text{F}\mu}m^{*}w/\sqrt{2e}\pi k_\text{B}T)^{2/3}$. The phase breaking time is known to saturate at low temperatures \cite{noguchi1996saturation} which is taken into consideration by fitting the equation $K=1/(1+A\cdot \exp(\tau_\text{t}/\tau_\varphi))\approx 1/(1+A\cdot \exp(B\cdot T^{2/3}))$ to the data. Here, $B=\tau_\text{t}/(\sqrt{E_\text{F}\mu}m^{*}w/\sqrt{2e}\pi k_\text{B})^{2/3}$ can be estimated from of the Fermi energy $E_\text{F}\approx 8.6\text{ meV}$, the effective mass $m^{*}\approx 0.062\text{ m}_0$ (from SdH measurements), $\tau_\text{t}\approx 11\text{ ps}$ and $\mu$ of the wafer material to be $B'\approx 0.14 \text{ K}^{-2/3}$. The best fit to the data is shown in Fig.~\ref{1Dchec} and yields $A=0.25$ and $B=0.11\text{ K}^{-2/3}$. The parameter $A$ is a measure for the highest value of $K_\text{max}=K(T_\text{bath}=0\text{ K})$ that is achievable in this quantum ring which is $K_\text{max}=1/(1+A)=0.8$.\\

\section*{V. Conclusion}
Thermal noise measurements are presented for Al${}_\text{x}$Ga${}_\text{1-x}$As/GaAs based 1D constrictions and asymmetric quantum rings. In the case of non-branched 1D waveguides the measured and calculated thermal noise fit each other as expected. However, in a branched network of electron waveguides, the measured noise exceeds the calculated one by up to 60 \%. It is proposed that this excess noise originates from the correlation of noise sources in the network which is mediated by the coherence of electrons in the quantum ring. Noise measurements at $T_\text{bath} = 4.2\text{ K}$ allow to estimate a correlation coefficient from the excess part of the noise. In general, the knowledge of the correlation coefficient of a quantum system could allow probing the phase coherence by noise measurements.

We gratefully acknowledge financial support by the priority programme “Nanostructured thermoelectrics” of the German Science Foundation DFG SPP 1386. We further thank Dr. R\"udiger Mitdank and Dr. Tobias Kramer for the fruitful scientific discussions.

\bibliography{bibliography} 

\begin{thebibliography}{14}%
\makeatletter
\providecommand \@ifxundefined [1]{%
 \@ifx{#1\undefined}
}%
\providecommand \@ifnum [1]{%
 \ifnum #1\expandafter \@firstoftwo
 \else \expandafter \@secondoftwo
 \fi
}%
\providecommand \@ifx [1]{%
 \ifx #1\expandafter \@firstoftwo
 \else \expandafter \@secondoftwo
 \fi
}%
\providecommand \natexlab [1]{#1}%
\providecommand \enquote  [1]{``#1''}%
\providecommand \bibnamefont  [1]{#1}%
\providecommand \bibfnamefont [1]{#1}%
\providecommand \citenamefont [1]{#1}%
\providecommand \href@noop [0]{\@secondoftwo}%
\providecommand \href [0]{\begingroup \@sanitize@url \@href}%
\providecommand \@href[1]{\@@startlink{#1}\@@href}%
\providecommand \@@href[1]{\endgroup#1\@@endlink}%
\providecommand \@sanitize@url [0]{\catcode `\\12\catcode `\$12\catcode
  `\&12\catcode `\#12\catcode `\^12\catcode `\_12\catcode `\%12\relax}%
\providecommand \@@startlink[1]{}%
\providecommand \@@endlink[0]{}%
\providecommand \url  [0]{\begingroup\@sanitize@url \@url }%
\providecommand \@url [1]{\endgroup\@href {#1}{\urlprefix }}%
\providecommand \urlprefix  [0]{URL }%
\providecommand \Eprint [0]{\href }%
\providecommand \doibase [0]{http://dx.doi.org/}%
\providecommand \selectlanguage [0]{\@gobble}%
\providecommand \bibinfo  [0]{\@secondoftwo}%
\providecommand \bibfield  [0]{\@secondoftwo}%
\providecommand \translation [1]{[#1]}%
\providecommand \BibitemOpen [0]{}%
\providecommand \bibitemStop [0]{}%
\providecommand \bibitemNoStop [0]{.\EOS\space}%
\providecommand \EOS [0]{\spacefactor3000\relax}%
\providecommand \BibitemShut  [1]{\csname bibitem#1\endcsname}%
\let\auto@bib@innerbib\@empty
\bibitem [{\citenamefont {Motchenbacher}\ and\ \citenamefont
  {Connelly}(1993)}]{motchenbacher1993low}%
  \BibitemOpen
  \bibfield  {author} {\bibinfo {author} {\bibfnamefont {C.~D.}\ \bibnamefont
  {Motchenbacher}}\ and\ \bibinfo {author} {\bibfnamefont {J.~A.}\ \bibnamefont
  {Connelly}},\ }\href@noop {} {\emph {\bibinfo {title} {Low noise electronic
  system design}}}\ (\bibinfo  {publisher} {Wiley},\ \bibinfo {year}
  {1993})\BibitemShut {NoStop}%
\bibitem [{\citenamefont {Nyquist}(1928)}]{nyquist1928thermal}%
  \BibitemOpen
  \bibfield  {author} {\bibinfo {author} {\bibfnamefont {H.}~\bibnamefont
  {Nyquist}},\ }\href@noop {} {\bibfield
  {journal} {\bibinfo  {journal} {Phys. Rev.}\ }\textbf {\bibinfo {volume}
  {\textbf{32}}},\ \bibinfo {pages} {110} (\bibinfo {year} {1928})}\BibitemShut
  {NoStop}%
\bibitem [{\citenamefont {Buchholz}\ \emph {et~al.}(2012)\citenamefont
  {Buchholz}, \citenamefont {Sternemann}, \citenamefont {Chiatti},
  \citenamefont {Reuter}, \citenamefont {Wieck},\ and\ \citenamefont
  {Fischer}}]{buchholz2012noise}%
  \BibitemOpen
  \bibfield  {author} {\bibinfo {author} {\bibfnamefont {S.~S.}\ \bibnamefont
  {Buchholz}}, \bibinfo {author} {\bibfnamefont {E.}~\bibnamefont
  {Sternemann}}, \bibinfo {author} {\bibfnamefont {O.}~\bibnamefont {Chiatti}},
  \bibinfo {author} {\bibfnamefont {D.}~\bibnamefont {Reuter}}, \bibinfo
  {author} {\bibfnamefont {A.~D.}\ \bibnamefont {Wieck}}, \ and\ \bibinfo
  {author} {\bibfnamefont {S.~F.}\ \bibnamefont {Fischer}},\ }
  \href@noop {} {\bibfield  {journal} {\bibinfo  {journal}
  {Phys. Rev. B}\ }\textbf {\bibinfo {volume} {\textbf{85}}},\ \bibinfo
  {pages} {235301} (\bibinfo {year} {2012})}\BibitemShut {NoStop}%
\bibitem [{\citenamefont {Van~der Ziel}(1986)}]{van1986noise}%
  \BibitemOpen
  \bibfield  {author} {\bibinfo {author} {\bibfnamefont {A.}~\bibnamefont
  {Van~der Ziel}},\ }\href@noop {} {\emph {\bibinfo {title} {Noise in solid
  state devices and circuits}}}\ (\bibinfo  {publisher} {Wiley-Interscience},\
  \bibinfo {year} {1986})\BibitemShut {NoStop}%
\bibitem [{\citenamefont {Blanter}\ and\ \citenamefont
  {B{\"u}ttiker}(2000)}]{blanter2000shot}%
  \BibitemOpen
  \bibfield  {author} {\bibinfo {author} {\bibfnamefont {Y.~M.}\ \bibnamefont
  {Blanter}}\ and\ \bibinfo {author} {\bibfnamefont {M.}~\bibnamefont
  {B{\"u}ttiker}},\ }\href@noop {} {\bibfield  {journal} {\bibinfo
  {journal} {Phys. Rep.}\ }\textbf {\bibinfo {volume} {\textbf{336}}},\
  \bibinfo {pages} {1--166} (\bibinfo {year} {2000})}\BibitemShut {NoStop}%
\bibitem [{\citenamefont {Kubo}(1966)}]{kubo1966fluctuation}%
  \BibitemOpen
  \bibfield  {author} {\bibinfo {author} {\bibfnamefont {R.}~\bibnamefont
  {Kubo}},\ }\href@noop {} {\bibfield  {journal}
  {\bibinfo  {journal} {Rep. Prog. Phys.}\ }\textbf {\bibinfo
  {volume} {\textbf{29}}},\ \bibinfo {pages} {255} (\bibinfo {year}
  {1966})}\BibitemShut {NoStop}%
\bibitem [{\citenamefont {Vasilescu}(2006)}]{vasilescu2006electronic}%
  \BibitemOpen
  \bibfield  {author} {\bibinfo {author} {\bibfnamefont {G.}~\bibnamefont
  {Vasilescu}},\ }\href@noop {} {\emph {\bibinfo {title} {Electronic noise and
  interfering signals: principles and applications}}}\ (\bibinfo  {publisher}
  {Springer Science \& Business Media},\ \bibinfo {year} {2006})\BibitemShut
  {NoStop}%
\bibitem [{\citenamefont {Jarrix}\ \emph {et~al.}(1997)\citenamefont {Jarrix},
  \citenamefont {Delseny}, \citenamefont {Pascal},\ and\ \citenamefont
  {Lecoy}}]{jarrix1997noise}%
  \BibitemOpen
  \bibfield  {author} {\bibinfo {author} {\bibfnamefont {S.}~\bibnamefont
  {Jarrix}}, \bibinfo {author} {\bibfnamefont {C.}~\bibnamefont {Delseny}},
  \bibinfo {author} {\bibfnamefont {F.}~\bibnamefont {Pascal}}, \ and\ \bibinfo
  {author} {\bibfnamefont {G.}~\bibnamefont {Lecoy}},\ }
  \href@noop {} {\bibfield  {journal} {\bibinfo  {journal}
  {J. Appl. Phys.}\ }\textbf {\bibinfo {volume} {\textbf{81}}},\
  \bibinfo {pages} {2651--2657} (\bibinfo {year} {1997})}\BibitemShut {NoStop}%
\bibitem [{\citenamefont {Zhang}\ and\ \citenamefont {Van
  Der~Ziel}(1984)}]{zhang1984correlation}%
  \BibitemOpen
  \bibfield  {author} {\bibinfo {author} {\bibfnamefont {X.}~\bibnamefont
  {Zhang}}\ and\ \bibinfo {author} {\bibfnamefont {A.}~\bibnamefont {Van
  Der~Ziel}},\ }\href@noop {} {\bibfield  {journal} {\bibinfo  {journal}
  {Physica B+ C}\ }\textbf {\bibinfo {volume} {\textbf{124}}},\ \bibinfo
  {pages} {62--64} (\bibinfo {year} {1984})}\BibitemShut {NoStop}%
\bibitem [{\citenamefont {Riha}\ \emph {et~al.}(2015)\citenamefont {Riha},
  \citenamefont {Miechowski}, \citenamefont {Buchholz}, \citenamefont
  {Chiatti}, \citenamefont {Wieck}, \citenamefont {Reuter},\ and\ \citenamefont
  {Fischer}}]{riha2015mode}%
  \BibitemOpen
  \bibfield  {author} {\bibinfo {author} {\bibfnamefont {C.}~\bibnamefont
  {Riha}}, \bibinfo {author} {\bibfnamefont {P.}~\bibnamefont {Miechowski}},
  \bibinfo {author} {\bibfnamefont {S.~S.}\ \bibnamefont {Buchholz}}, \bibinfo
  {author} {\bibfnamefont {O.}~\bibnamefont {Chiatti}}, \bibinfo {author}
  {\bibfnamefont {A.~D.}\ \bibnamefont {Wieck}}, \bibinfo {author}
  {\bibfnamefont {D.}~\bibnamefont {Reuter}}, \ and\ \bibinfo {author}
  {\bibfnamefont {S.~F.}\ \bibnamefont {Fischer}},\ }\href@noop {} {\bibfield  {journal}
  {\bibinfo  {journal} {Appl. Phys. Lett.}\ }\textbf {\bibinfo {volume}
  {\textbf{106}}},\ \bibinfo {pages} {083102} (\bibinfo {year}
  {2015})}\BibitemShut {NoStop}%
\bibitem [{\citenamefont {van~der Ziel}, \citenamefont {Zhang},\ and\
  \citenamefont {Pawlikiewicz}(1986)}]{van1986location}%
  \BibitemOpen
  \bibfield  {author} {\bibinfo {author} {\bibfnamefont {A.}~\bibnamefont
  {van~der Ziel}}, \bibinfo {author} {\bibfnamefont {X.}~\bibnamefont {Zhang}},
  \ and\ \bibinfo {author} {\bibfnamefont {A.~H.}\ \bibnamefont
  {Pawlikiewicz}},\ }\href@noop {}
  {\bibfield  {journal} {\bibinfo  {journal} {IEEE Tran. Electron
  Devices}\ }\textbf {\bibinfo {volume} {\textbf{33}}},\ \bibinfo {pages}
  {1371--1376} (\bibinfo {year} {1986})}\BibitemShut {NoStop}%
\bibitem [{\citenamefont {Hansen}\ \emph {et~al.}(2001)\citenamefont {Hansen},
  \citenamefont {Kristensen}, \citenamefont {Pedersen}, \citenamefont
  {S{\o}rensen},\ and\ \citenamefont {Lindelof}}]{hansen2001mesoscopic}%
  \BibitemOpen
  \bibfield  {author} {\bibinfo {author} {\bibfnamefont {A.~E.}\ \bibnamefont
  {Hansen}}, \bibinfo {author} {\bibfnamefont {A.}~\bibnamefont {Kristensen}},
  \bibinfo {author} {\bibfnamefont {S.}~\bibnamefont {Pedersen}}, \bibinfo
  {author} {\bibfnamefont {C.}~\bibnamefont {S{\o}rensen}}, \ and\ \bibinfo
  {author} {\bibfnamefont {P.}~\bibnamefont {Lindelof}},\ }
  \href@noop {} {\bibfield  {journal} {\bibinfo  {journal} {Phys.
  Rev. B}\ }\textbf {\bibinfo {volume} {\textbf{64}}},\ \bibinfo {pages}
  {045327} (\bibinfo {year} {2001})}\BibitemShut {NoStop}%
\bibitem [{\citenamefont {Seelig}\ and\ \citenamefont
  {B{\"u}ttiker}(2001)}]{seelig2001charge}%
  \BibitemOpen
  \bibfield  {author} {\bibinfo {author} {\bibfnamefont {G.}~\bibnamefont
  {Seelig}}\ and\ \bibinfo {author} {\bibfnamefont {M.}~\bibnamefont
  {B{\"u}ttiker}},\ }\href@noop {} {\bibfield  {journal} {\bibinfo  {journal}
  {Phys. Rev. B}\ }\textbf {\bibinfo {volume} {64}},\ \bibinfo {pages}
  {245313} (\bibinfo {year} {2001})}\BibitemShut {NoStop}%
\bibitem [{\citenamefont {Noguchi}\ \emph {et~al.}(1996)\citenamefont
  {Noguchi}, \citenamefont {Ikoma}, \citenamefont {Odagiri}, \citenamefont
  {Sakakibara},\ and\ \citenamefont {Wang}}]{noguchi1996saturation}%
  \BibitemOpen
  \bibfield  {author} {\bibinfo {author} {\bibfnamefont {M.}~\bibnamefont
  {Noguchi}}, \bibinfo {author} {\bibfnamefont {T.}~\bibnamefont {Ikoma}},
  \bibinfo {author} {\bibfnamefont {T.}~\bibnamefont {Odagiri}}, \bibinfo
  {author} {\bibfnamefont {H.}~\bibnamefont {Sakakibara}}, \ and\ \bibinfo
  {author} {\bibfnamefont {S.~N.}\ \bibnamefont {Wang}},\ }
  \href@noop {} {\bibfield  {journal} {\bibinfo
  {journal} {J. Appl. Phys.}\ }\textbf {\bibinfo {volume} {80}},\
  \bibinfo {pages} {5138--5144} (\bibinfo {year} {1996})}\BibitemShut {NoStop}%
\end{thebibliography}%

\end{document}